\documentclass[a4paper]{article}

\usepackage[english]{babel}
\usepackage[utf8x]{inputenc}
\usepackage[T1]{fontenc}
\usepackage{ upgreek }
\usepackage{gensymb} 
\usepackage{threeparttable}
\usepackage{tabularx}
\usepackage{ntheorem}
\newtheorem{theorem}{Theorem}

\usepackage{cite}
\usepackage{mathrsfs}
\usepackage{pifont}
\usepackage{amssymb}
\usepackage{array, caption}
\usepackage{amsmath}
\usepackage{booktabs}
\makeatletter
\newcommand{\ssymbol}[1]{^{\@fnsymbol{#1}}}
\makeatother

\usepackage{amsmath}
\usepackage{graphicx}
\usepackage[colorinlistoftodos]{todonotes}
\usepackage[colorlinks=true, allcolors=blue]{hyperref}

\usepackage[a4paper,top=2.5cm,bottom=2cm,left=3cm,right=3cm,marginparwidth=1.75cm]{geometry}

\usepackage{authblk}
\title{ Semi-quantum private comparison and its generalization to the key agreement, summation, and anonymous ranking}
\author[1]{Chong-Qiang Ye}
\author[2]{Jian Li\thanks{Corredponding author:  Lijian@bupt.edu.cn}}
\author[2]{Xiu-Bo Chen}
\author[1]{Yanyan Hou}
\author[1]{Zhou Wang}
\affil[1]{ School of Artificial Intelligence, Beijing University of Posts Telecommunications, Beijing 100876, China.}
\affil[2]{Information Security Center, State Key Laboratory of Networking and Switching Technology,
Beijing University of Posts and Telecommunications, Beijing 100876, China}

\begin{document}
\date{}
  \maketitle

\begin{abstract}

Semi-quantum protocols construct connections between quantum users and ``classical'' users who can only perform certain ``classical'' operations. In this paper, we present a new semi-quantum private comparison protocol based on entangled states and single particles, which does not require pre-shared keys between the ``classical'' users to guarantee the security of their private data. By utilizing multi-particle entangled states and single particles, our protocol can be easily extended to multi-party scenarios to meet the requirements of multiple ``classical'' users who want to compare their private data. The security analysis shows that the protocol can effectively prevent attacks from outside eavesdroppers and adversarial participants. Besides, we generalize the proposed protocol to other semi-quantum protocols such as semi-quantum key agreement, semi-quantum summation, and semi-quantum anonymous ranking protocols. We compare and discuss the proposed protocols with previous similar protocols. The results show that our protocols satisfy the demands of their respective counterparts separately. Therefore, our protocols have a wide range of application scenarios.

\end{abstract}

\section{Introduction}
\label{intro}
Quantum cryptography \cite{1,2,3} has made tremendous progress in the last few decades, because its security is based on quantum laws rather than computational complexity. Using quantum resources to solve problems in classical cryptography has become a research hotspot. Many quantum cryptography protocols have been proposed, including quantum key distribution (QKD)\cite{4,5}, quantum secure direct communication (QSDC)\cite{6,7,8}, quantum secret sharing (QSS)\cite{10,11}, and so on.

Quantum private comparison (QPC) \cite{12} is one of the essential applications of quantum cryptography, whose goal is to allow $n(n\ge2)$ users to compare whether their private data are the same without disclosing their data. Comparing two or more data to determine if they are identical has many applications in information science, such as malware detection and bug search\cite{13}. As a result, QPC has attracted much attention from academia because it can provide quantum mechanics-based security. Generally, the QPC protocol should meet the following three conditions: (1) Correctness: All users should obtain the comparison result simultaneously, and the result should be correct. (2) Privacy: Each user's data is confidential. (3) Security: Any attack will be detected, and attackers cannot steal users' data without being detected. In addition, since it is impossible to construct a safe equation function in a two-party scenario\cite{14}, some additional assumptions are always needed in QPC, such as a third party (TP), who may be adversarial. Specifically, TP is fully compliant with the protocol requirements, but otherwise, she can do whatever she likes, including stealing the users' data in all possible ways. Up to now, many QPC protocols have been designed for different quantum states such as single particle states, multi-particle entangled states, and multilevel quantum systems \cite{15,16,17,18,19,20,21}. It is easy to notice that all the above QPC protocols \cite{15,16,17,18,19,20,21} require all participants to be quantum capable in order to ensure the security of the protocol. However, not all parties can afford the high cost of quantum devices under the current technological conditions. A natural question that is raised is how much quantumness is required to achieve unconditional security? Alternatively, do all users involved in a cryptographic scheme need to have the ability to prepare and measure arbitrary quantum states?

Fortunately, Boyer et al. \cite{22,23} answered this question. They proposed a semi-quantum key distribution (SQKD) protocol and constructed the first semi-quantum model. In the semi-quantum model, the ``quantum'' user has full quantum power while the ``classical'' user's quantum power is limited. The ``classical'' user can only carry out the computational base $\{|0\rangle,|1\rangle\}$ preparation and measurement, direct reflection, and reordering operations. It provides a theoretical solution for the high cost of quantum devices, and then semi-quantum has gradually become a research hotspot. Many semi-quantum-based studies have been proposed, such as semi-quantum key distribution (SQKD) \cite{22,23,24,25,26}, semi-quantum secure direct communication (SQSDC) \cite{27,28,29}, semi-quantum secret sharing (SQSS) \cite{30,31,32}, and so on.

Semi-quantum private comparison (SQPC) was first put forward by Chou et al. \cite{33} in 2016, where two classical users, Alice and Bob, want to know whether their private data are the same with the help of TP who has full quantum capability and may be adversarial. This imbalance of participants' abilities makes the protocol more interesting and attracts the attention of many scholars. Then several SQPC protocols based on different quantum states were proposed. For example, Thapliyala et al. \cite{34} put forward an SQPC protocol based on Bell states. Ye et al. \cite{35} utilized two-particle product states to implement SQPC, in which two classical users need to use the SQKD protocol\cite{25} to generate the pre-shared keys for encrypting their private data. In 2019, Lin et al. \cite{36} proposed an SQPC protocol based on single-particle states instead of entangled states. In 2021, Yan et al. \cite{37} designed an SQPC protocol using three-particle G-like states, in which pre-shared keys are also required. In the same year, Ye et al.\cite{38} proposed an SQPC protocol based on circular transmission. Unlike the previous protocols, in this protocol, the qubits travel from TP to Alice, to Bob, then back to TP. It's well known that pre-shared keys play an essential role in SQPC protocols. Many protocols rely on them to enable key sharing between two classical users and ensure the protocol's security. However, this requires the consumption of additional qubits, which significantly reduces the efficiency of the protocol. In addition, most previous SQPC protocols are only applicable to two-party scenarios. If more than two classical users want to compare their private data, previous protocols have many limitations. For example, if a two-party SQPC protocol is used to solve a multi-party equality comparison problem, multiple executions are required, which increases the time complexity and reduces the efficiency of the entire process. Therefore, how to improve the efficiency of SQPC protocol and enrich the application scenarios is gradually becoming a hot research topic.

In this work, we present a new SQPC protocol based on entangled states and single particles, where pre-shared keys are not required, thus avoiding protocol inefficiency due to additional SQKD protocol between classical users. By exploiting multi-particle entangled states, our protocol can be easily extended to multi-party scenarios. It can be used to solve the problem of multiple classical users' privacy data comparisons. For the security of the protocol, we prove that the attacks from the outside eavesdroppers and the attacks from the adversarial users and TP are invalid. In our protocol, TP can establish a secure key with each classical user, and each classical user can also create a secure key for each other. Following this fact, we generalize the proposed SQPC protocol to the semi-quantum key agreement (SQKA), semi-quantum summation (SQS), and semi-quantum anonymous ranking (SQAR). To better highlight the characteristics of these protocols, we compare the proposed protocols with their respective counterparts separately. The detailed comparison results are shown in Tables 3, 4, and 5.

The remaining organization is shown below. Sect. 2 describes the detailed steps of the proposed SQPC and extends the protocol to multi-party scenarios. Sect. 3 analyzes the security of the proposed protocol. In Sect. 4, we generalize the proposed SQPC protocol to other semi-quantum protocols, including SQKA, SQS, and SQAR protocols. Then, in Sect. 5, we compare the proposed protocols with previous similar studies. Finally, this paper concludes in Sect. 6.


\section{ The proposed SQPC protocol}

In our protocol, two classical users, Alice and Bob, are limited to the following operations: (1) \textbf{\emph{measure}}: measure the qubit in $Z$ basis $\{|0\rangle,|1\rangle\}$ and regenerate one in the same state (e.g., $|0\rangle \rightarrow |0\rangle$, $|1\rangle \rightarrow |1\rangle$). (2) \textbf{\emph{reflect}}: reflect the qubit directly. (3) \textbf{\emph{reorder}}: reorder the qubit via delay lines. They want to compare their private data with the help of quantum TP, who has the full quantum capability and may be adversarial. Alice's private binary string is denoted as $M_A=[m^{1}_{A},m^{2}_{A},\dots,m^{n}_{A}]$, while Bob's private binary string is denoted as  $M_B=[m^{1}_{B},m^{2}_{B},\dots,m^{n}_{B}]$. The detailed protocol steps are as follows.

\textbf{Step 1:} TP prepares $4n$ Bell states all in the state of $|\phi^+\rangle=\frac{1}{\sqrt 2}(|00\rangle+|11\rangle)_{AB}$, and divides the first and second qubits of these Bell states into two sequences: 
\begin{equation}
\begin{split}
\label{Bell-sequence}
&S_A:  P^{1}_{A},P^{2}_{A},\dots, P^{4n}_{A},\quad S_B:  P^{1}_{B},P^{2}_{B},\dots, P^{4n}_{B}.
\end{split}
\end{equation}	
Then, TP prepares the other two sequences: 
\begin{equation}
\begin{split}
\label{Z-sequence-1}
&T_A: P^{1}_{TA},P^{2}_{TA},\dots, P^{4n}_{TA},\quad T_B: P^{1}_{TB},P^{2}_{TB},\dots, P^{4n}_{TB},
\end{split}
\end{equation}	
where each qubit is randomly from the set $\{|0\rangle, |1\rangle\}$.

\textbf {Step 2:}  TP randomly inserts $T_A$ ($T_B$) into $S_A$ ($S_B$) to form a new sequence $S^{*}_{A}$ ($S^{*}_{B}$). Subsequently, TP sends $S^{*}_{A}$ and $S^{*}_{B}$ to Alice and Bob, respectively.

\textbf {Step 3:} For each received qubit, Alice (Bob) randomly chooses the \textbf{\emph{measure}} or \textbf{\emph{reflect}} operation. Alice (Bob) will record the measurement result, when she (he) chooses the \textbf{\emph{measure}} operation.

\textbf {Step 4:} After TP receives all the qubits, she divides the qubits of $T_A$ ($T_B$) from  qubits of $S_A$ ($S_B$) and performs different measurements. In more detail, TP measures the qubits of $T_A$ and $T_B$ in the $Z$ basis, while for qubits of $S_A$ and $S_B$ (i.e., $P^{i}_{A}$ and $P^{i}_{B}$, $i\in\{1,2,\dots,4n\}$) she will perform Bell state measurements. Then, TP publishes her measurement results of $S_A$ and $S_B$, and the positions of these qubits. Note that the measurement results of $T_A$ and $T_B$ are kept in her hands and not leaked to Alice and Bob (They will be discussed later).

\textbf {Step 5:} Alice and Bob will discuss eavesdropping and TP's honesty in this step. According to Alice and Bob's operations and the information provided by TP, the following three cases will happen ( It is expected that there are $n$ qubits in Case 1.):
\begin{itemize}
\item \textbf{Case 1}: When Alice and Bob choose the \textbf{\emph{measure}} operation on the qubits $P^{i}_{A}$ and $P^{i}_{B}$, TP will publish the result as either $|\phi^+\rangle$ or $|\phi^-\rangle$. In this case, Alice and Bob's measurement result should be the same, while TP knows nothing about their measurements because TP performs Bell state measurements, not $Z$-basis measurements, on the qubits $P^{i}_{A}$ and $P^{i}_{B}$. Thus, Alice and Bob can estalish a secure key sequence denoted as $K_{AB}=[k^{1}_{AB},k^{2}_{AB},\dots,k^{n}_{AB}]$.

\item \textbf{Case 2}: When they choose the \textbf{\emph{reflect}} operatin on the qubits $P^{i}_{A}$ and $P^{i}_{B}$, TP should always publish $|\phi^+\rangle$. This case is used for checking the honesty of TP and eavesdropping. If error rate surpasses the threshold, the protocol ends.

\item \textbf{Case 3}: When they perform different operations, this case will be discarded.

\end{itemize}

\textbf {Step 6:} After eavesdropping checking, Alice and Bob tell TP their operations on the qubits of $T_A$ and $T_B$. Note that in the sequence $T_A$ ($T_B$), there are $2n$ qubits performed the \textbf{\emph{measure}} operation and $2n$ qubits performed the \textbf{\emph{reflect}} operation. As a result, two scenarios can arise:

1) For the qubits performed the \textbf{\emph{reflect}} operation, TP compares these qubits' initial states and measurement results recored in step 4. If the error rate exceeds the threshold, the protocol terminates and restarts.

2) For the qubits performed the \textbf{\emph{measure}} operation, Alice (Bob) picks out $n$ qubits used for eavesdropping checking again. If there is no Eve online, for the selected qubits, Alice's (Bob's) measurements, TP's measurements, and the initial state of these qubits, all three are the same. For the remaining $n$ qubits, Alice (Bob) and TP can establish a secure key sequence because TP also measured them in step 4 with the $Z$-basis, while Bob (Alice) knows nothing about them. We use $K_{TA}=[k^{1}_{TA},k^{2}_{TA},\dots,k^{n}_{TA}]$ ($K_{TB}=[k^{1}_{TB},k^{2}_{TB},\dots,k^{n}_{TB}]$) to represent the secret key between Alice and TP (Bob and TP).

\textbf {Step 7:} Alice and Bob calculate $Q^{j}_{A}=k^{j}_{AB}\oplus k^{j}_{TA}\oplus m^{j}_{A}$ and $Q^{j}_{B}=k^{j}_{AB}\oplus k^{j}_{TB}\oplus m^{j}_{B}$, respectively, where $\oplus$ is the modulo 2 summation, and $j\in\{1,2,\dots,n\}$. Afterwards, Alice and Bob send $Q_A=[Q^{1}_{A},Q^{2}_{A},\dots,Q^{n}_{A}]$ and $Q_B=[Q^{1}_{B},Q^{2}_{B},\dots,Q^{n}_{B}]$ to TP.

\textbf {Step 8:} TP calculates $R^{j}=Q^{j}_{A}\oplus Q^{j}_{B}\oplus k^{j}_{TA}\oplus k^{j}_{TB}$. If $R^j=0$ for $j=1,2,\dots,n$, she will conclude that the secrets of Alice and Bob are equal. Otherwise, their secrets are not equal. Finally, TP announces the comparison result to Alice and Bob.

\subsection {Correctness of protocol}
In our protocol, Alice and Bob can establish the secure key sequence $K_{AB}=[k^{1}_{AB},k^{2}_{AB},\dots,k^{n}_{AB}]$ based on $S_A$ and $S_B$. Then, TP can establish the secure key sequence $K_{TA}=[k^{1}_{TA},k^{2}_{TA},\dots,k^{n}_{TA}]$ ($K_{TB}=[k^{1}_{TB},k^{2}_{TB},\dots,k^{n}_{TB}]$) with Alice (Bob) based on the measurement result of $T_A$ ($T_B$). Afterwards, Alice uses $K^{j}_{AB}$ and $K^{j}_{TA}$ to encrypt her secret information $m^j_A$ as 

\begin{equation}
Q^{j}_{A}=k^{j}_{AB}\oplus k^{j}_{TA}\oplus m^{j}_{A}. 
\end{equation}
Bob also uses $K^{j}_{AB}$ and $K^{j}_{TB}$ to encrypt his secret information $m^j_B$ as 

\begin{equation}
Q^{j}_{B}=k^{j}_{AB}\oplus k^{j}_{TB}\oplus m^{j}_{B}. 
\end{equation}
Finally, TP calculates $R^{j}=Q^{j}_{A}\oplus Q^{j}_{B}\oplus k^{j}_{TA}\oplus k^{j}_{TB}$. It is easy to obtain that 
\begin{equation}
\begin{aligned}
R^{j}&=Q^{j}_{A}\oplus Q^{j}_{B}\oplus k^{j}_{TA}\oplus k^{j}_{TB}\\
&=k^{j}_{AB}\oplus k^{j}_{TA}\oplus m^{j}_{A}\oplus k^{j}_{AB}\oplus k^{j}_{TB}\oplus m^{j}_{B}\oplus k^{j}_{TA}\oplus k^{j}_{TB}\\
&=m^{j}_{A}\oplus m^{j}_{B}.
\end{aligned}
\end{equation}
If $R^j=0$ for $j=1,2,\dots,n$, she will conclude that the secrets of Alice and Bob are equal. Otherwise, their secrets are not equal. The results show that the proposed protocol can guarantee the correctness of the output.

\subsection {Extend to multi-party SQPC protocol}

Our protocol can be extended into a multi-party SQPC protocol involving $L (L>2)$ ``classical'' users. We use $C_1,C_2,\cdots,C_L$ to denote these  ``classical'' users. Only TP is a fully quantum user in this protocol, while other users are all ``classical'' whose quantum power is limited. Here, $C_l$ ($l=1,2,\dots,L$) has a private binary string $M_{C_l}=[m^{1}_{C_l},m^{2}_{C_l},\dots,m^{n}_{C_l}]$. 

We use the $L$-particle GHZ entangled state $|\Psi^+\rangle_L=\frac{1}{\sqrt 2}(|0\rangle^{\otimes L}+|1\rangle^{\otimes L})$  and single particles $\{|0\rangle,|1\rangle\}$ as initial quantum resourcce to achieve multi-party SQPC. The specific steps are as follows.

\textbf{Step 1*:} TP prepares $2^Ln$  $L$-particle GHZ entangled states all in the form of  $|\Psi^+\rangle$. Then, she divides these GHZ states into $L$ sequences 
\begin{equation}
\begin{split}
\label{GHZ-sequence}
&S_{C_1}:  P^{1}_{C_1},P^{2}_{C_1},\dots,P^{2^Ln}_{C_1},\\
&  \qquad \vdots\\
&S_{C_l}:  P^{1}_{C_l},P^{2}_{C_l},\dots,P^{2^Ln}_{C_l},\\
&  \qquad \vdots\\
&S_{C_L}:  P^{1}_{C_L},P^{2}_{C_L},\dots,P^{2^Ln}_{C_L}.
\end{split}
\end{equation}	
TP prepares the other $L$ sequences 
\begin{equation}
\begin{split}
\label{Z-sequence}
&T_{C_1}:  P^{1}_{TC_1},P^{2}_{TC_1},\dots,P^{4n}_{TC_1},\\
&  \qquad \vdots\\
&T_{C_l}:  P^{1}_{TC_l},P^{2}_{TC_l},\dots,P^{4n}_{TC_l},\\
&  \qquad \vdots\\
&T_{C_L}:  P^{1}_{TC_L},P^{2}_{TC_L},\dots,P^{4n}_{TC_L},
\end{split}
\end{equation}	
where each qubit is randomly from the set $\{|0\rangle, |1\rangle\}$.

\textbf {Step 2*:}  For $l=1,2,\dots,L$, TP randomly inserts $T_{C_l}$ into $S_{C_l}$ to form a new sequence $S^{*}_{C_l}$. Subsequently, TP sends $S^{*}_{C_l}$ to the classical user $C_l$.

\textbf {Step 3*:} For each received qubit, $C_l$ randomly chooses the \textbf{\emph{measure}} or \textbf{\emph{reflect}} operation. $C_l$ will record the measurement result, when he chooses the \textbf{\emph{measure}} operation. 		 			
		 			
\textbf {Step 4*:} After TP receives all the qubits, she divides the qubits of $T_{C_l}$ from  qubits of $S_{C_l}$ and performs different measurements. In more detail, TP measures the qubits of $T_{C_l}$ in the $Z$ basis, while for qubits of $S_{C_1},S_{C_2},\dots,S_{C_L}$ (i.e., $P^{i}_{C_1},P^{i}_{C_2},\dots,P^{i}_{C_L}$, $i\in\{1,2,\dots,2^L n\}$, which come from the same GHZ state), she will perform $L$-particle GHZ state measurements. Then, TP publishes her measurement results of $S_{C_1},S_{C_2},\dots,S_{C_L}$ in the form of $L$-particle GHZ state and the positions of these qubits. Note that the measurement results of $T_{C_l}$ are kept in her hands and not leaked to $C_l$ .

\textbf {Step 5*:} $C_1,C_2,\cdots,C_L$ discuss eavesdropping and TP's honesty in this step. According to their operations and the information provided by TP, the following three cases will happen ( It is expected that there are $n$ qubits in Case b.):

\begin{itemize}
\item [\textbf{a)}] If all $C_1,C_2,\cdots,C_L$ choose the \textbf{\emph{reflect}} operation on the qubits $P^{i}_{C_1},P^{i}_{C_2},\dots,P^{i}_{C_L}$, TP should always get $|\Psi^+\rangle$. This case is used for checking TP's honesty and eavesdropping. When the error rate of this case surpasses the threshold, the protocol ends.

\item [\textbf{b)}] If all $C_1,C_2,\cdots,C_L$ choose the \textbf{\emph{measure}} operation on the qubits $P^{i}_{C_1},P^{i}_{C_2},\dots,P^{i}_{C_L}$, respectively, they will have the same measurement results, recored as $K_C=[k^{1}_{C},k^{2}_{C},\dots,k^{n}_{C}]$. While TP knows nothing about their measurements because TP performs $L$-particle GHZ state measurements, not $Z$-basis measurements, on the qubits $P^{i}_{C_1},P^{i}_{C_2},\dots,P^{i}_{C_L}$.

\item [\textbf{c)}] If their operations are inconsistent, $C_1,C_2,\cdots,C_L$ will discard these qubits.
\end{itemize}

\textbf {Step 6*:} After eavesdropping checking, $C_l$ ($l=1,2,\dots,L$) tells TP his operations on the qubits of $T_{C_l}$. Note that in the sequence $T_{C_l}$, there are $2n$ qubits performed the \textbf{\emph{measure}} operation and $2n$ qubits performed the \textbf{\emph{reflect}} operation. As a result, two scenarios can arise:

i) For the qubits performed the \textbf{\emph{reflect}} operation, TP compares these qubits' initial states and measurement results recored in step 4*. The error rate exceeding the threshold will lead to the termination and restart of the protocol.
 
 ii) For the qubits performed the \textbf{\emph{measure}} operation, $C_l$ picks out $n$ qubits used for eavesdropping checking again. If there is no Eve online, for the selected qubits, $C_l$'s measurements, TP's measurements, and the initial state of these qubits, all three are the same. For the remaining $n$ qubits, $C_l$ and TP can establish a secure key sequence because TP also measured them in step 4* with the $Z$-basis. We use $K_{TC_l}=[k^{1}_{TC_l},k^{2}_{TC_l},\dots,k^{n}_{TC_l}]$ to represent the secret key between $C_l$ and TP.

\textbf {Step 7*:} For the arbitrary two classical users $C_l$ and $C_g$ ($l,g=1,2,\dots,L$ and $l\neq g$), they can encrypt their secret as  $Q^{j}_{C_l}=k^{j}_{C}\oplus k^{j}_{TC_l}\oplus m^{j}_{C_l}$ and $Q^{j}_{C_g}=k^{j}_{C}\oplus k^{j}_{TC_g}\oplus m^{j}_{C_g}$, respectively, where $\oplus$ is the modulo 2 summation, and $j\in\{1,2,\dots,n\}$.  Afterwards, $C_l$ and $C_g$ publish $Q_{C_l}=[Q^{1}_{C_l},Q^{2}_{C_l},\dots,Q^{n}_{C_l}]$ and $Q_{C_g}=[Q^{1}_{C_g},Q^{2}_{C_g},\dots,Q^{n}_{C_g}]$ to TP via a public channel.

\textbf {Step 8*:} TP calculates $R^{j}=Q^{j}_{C_l}\oplus Q^{j}_{C_g}\oplus k^{j}_{TC_l}\oplus k^{j}_{TC_g}$. If $R^j=0$ for $j=1,2,\dots,n$, she will conclude that the secrets of $C_l$ and $C_g$ are equal. Otherwise, their secrets are not equal. Finally, TP announces the comparison result to $C_l$ and $C_g$.

It can be concluded that our protocol can be successfully used for multi-party SQPC, and enables arbitrary pair's comparison of equality among $n$ classical users.

\section{Security analysis}

In the above section, we gave the specific steps of the SQPC protocol and realized the privacy comparison under the scenarios of two classical and multiple classical users, respectively. Let us now analyze the security of the proposed protocol. Here, we would like to investigate the security of the proposed two-party SQPC protocol since the security of the proposed multi-party SQPC protocol can be analyzed similarly.
 In more detail, we will show that the attacks from outside eavesdroppers (i.e., the external attack) and the attacks from adversarial users and TP (i.e., the internal attack) are invalid.

\subsection{External attack}

In what follows, we show that Eve will be captured regardless of the attacks she uses, such as the intercept-resend attack, the measure-resend attack, the entangle-measure attack, the Double-CNOT attack and the Trojan horse attack. Although not all categories of attacks are covered, most of typical attacks usually analyzed in SQPC protocols are considered.

\subsubsection{The intercept-resend attack}
Suppose an external eavesdropper, Eve, uses the intercept-resend attack to obtain some helpful information. She may first intercept all the qubits sent from TP to Alice (Bob) and return them to TP directly. Then, she generates fake qubits in basis $ \{|0\rangle, |1\rangle \}$, and sends these fake qubits to Alice (Bob). After Alice (Bob) performs the operations on these qubits, Eve catches these qubits from Alice (Bob) and measures them in basis $ \{|0\rangle, |1\rangle \}$ to obtain some helpful information. 

However, Eve's attack is invalid because Eve cannot distinguish which qubits are performed \textbf{\emph{measure}} operation and which qubits are performed  \textbf{\emph{reflect}} operation. In step 4, TP will publish her measurement results of $S_A$ and $S_B$ in the form of Bell state. After Eve's intercept-resend attack, TP's measurement results should always be $|\phi^+\rangle$, since Eve intercepts the qubits of TP and returns them directly. This unusual situation would be judged as the presence of an eavesdropper in the quantum channel because, without Eve's attack, the TP's measurement results must be $|\phi^+\rangle$ only if both classical users \textbf{\emph{reflect}} qubits, while in all other cases $|\phi^+\rangle$ and $|\phi^-\rangle$ appear randomly. In addition, Eve's fake qubits are not the same as the original ones. Alice and Bob will discover this attack when they perform eavesdropping detection in step 6. Therefore, Eve's intercept-resend attack will not succeed.

\subsubsection{The measure-resend attack}
In our protocol, Eve may launch the measure-resend attack to steal some useful information. To obtain the information of Alice and Bob, Eve will use Z-basis to measure the intercepted qubits, since Alice and Bob only work on the Z-basis. Specifically, Eve first intercepts and measures all the qubits sent by TP. Then she sends the measured qubits to Alice (Bob) directly. After that, Alice (Bob) performs the \textbf{\emph{measure}} operation or the \textbf{\emph{reflect}} operation on the received qubits. Subsequently, Eve intercepts the qubits sent from Alice (Bob) to TP, and measures these qubits to extract some useful information. 

However, through this kind of attack, Eve will inevitably be detected in step 5, as her measurement may destroy $|\phi^+\rangle$. More precisely, if Alice (Bob) chooses the \textbf{\emph{measure}} operation, Eve's measure-resend attack induces no errors. If Alice and Bob both choose the \textbf{\emph{reflect}} operation on the qubits $P^{i}_{A}$ and $P^{i}_{B}$, Eve's attack will destroy the state of $|\phi^+\rangle$, and TP will have a $\frac{1}{2}$ probability to obtain the wrong measurement result $|\phi^-\rangle$. Therefore, adopting the measure-resend attack, Eve will unavoidably be found by Alice and Bob.

\subsubsection{The entangle-measure attack}
The proposed protocol uses the two-way quantum channels, so the entangle-measure attack can be modeled as two unitaries $U_E$ and $U_F$. Here, $U_E$ is the attack operator applied on the qubits sent from TP to Alice (Bob) while $U_F$ is the attack operator applied on the qubits sent from Alice (Bob) to TP. Note that $U_E$ and $U_F$ share a common probe space with initial state $|0\rangle_E$. In this kind of attack, Eve may perform $(U_E,U_F)$ on the target qubit and her probe $|0\rangle_E$ to extract some useful information. In the following, we will demonstrate that Eve is unable to access Alice and Bob's information without being discovered.

\begin{theorem}
Asumme that Eve launch attack $(U_E,U_F)$ on the target qubit and her probe $|0\rangle_E$. For this attack inducing no error, the final state of Eve's probe should be independent of the target qubit. That is, Eve obtains nothing about the secret information of Alice and Bob.
\end{theorem}

\noindent \textbf{Proof.} In our protocol, we use the Bell states $|\phi^+\rangle$ and single particles $\{|0\rangle,|1\rangle\}$ as initial quantum resourcce. We first analyze the attack $(U_E,U_F)$ on the qubits $|0\rangle$ and $|1\rangle$ (The analysis of the Bell state will be discussed later). 

\textbf{\emph{(1) Eve performs $(U_E,U_F)$ on the single particles}}

Eve first launches $U_E$ on the qubits from TP to Alice (Bob). The effect of $U_E$ on the qubits $|0\rangle$ and $|1\rangle$ is:
\begin{equation}
\begin{aligned}
U_E|0,0\rangle_{TE} =\alpha|0,E_0\rangle + \beta|1,E_1\rangle,\\
U_E|1,0\rangle_{TE} =\nu|0,E_2\rangle + \mu|1,E_3\rangle,
\end{aligned}
\end{equation}
where $T$ and $E$ represent the target qubit and Eve's probe, respectively. $|E_0\rangle$, $|E_1\rangle$, $|E_2\rangle$, $|E_3\rangle$ are the pure states determined by $U_E$, and $\|\alpha^2\|+\|\beta^2\|=1$, $\|\nu^2\|+\|\mu^2\|=1$.


Then, Alice (Bob) receives the qubits from TP, she (he) will choose either to \textbf{\emph{reflect}} or \textbf{\emph{measure}} the qubits.
Here, we fouces on the qubits that Alice (Bob) chooses the \textbf{\emph{measure}} operation. After Alice (Bob) chooses the \textbf{\emph{measure}} operation, Alice's (Bob's) measurement results should be the same as the initial state of the qubits sent by TP. If Eve wants to be undetected, $U_E$ needs to satisfy the conditions: $\beta=\nu=0$ and $\alpha=\mu=1$. Thus, the Eq. (8) can be reduced to 
\begin{equation}
\begin{aligned}
U_E|0,0\rangle_{TE} =|0,E_0\rangle,\quad U_E|1,0\rangle_{TE} =|1,E_3\rangle.
\end{aligned}
\end{equation}

After Alice's (Bob's) operations, Eve will perform $U_F$ on the qubits from Alice (Bob) to TP. Recall the step 6, for the qubits performed the \textbf{\emph{measure}} operation, Alice's (Bob's) measurement results should also be the same as TP's measurement results. Therefore, for Eve to be undetected in step 6, $U_F$ must meet the following conditions:
\begin{equation}
\begin{aligned}
U_F|0,E_0\rangle=|0,F_0\rangle,\quad  U_F(|1,E_1\rangle)=|1,F_1\rangle, \\
U_F|0,E_2\rangle=|0,F_2\rangle,\quad  U_F(|1,E_3\rangle)=|1,F_3\rangle, \\
\end{aligned}
\end{equation}
which means that $U_F$ cannot alter the states of the qubits from Alice (Bob) to TP.
 Putting everyting together, if Eve wants to induce no error when she perfroms $(U_E,U_F)$ on the single particles, the Eq. (9) and Eq. (10) must simultaneously hold. Thus, after Eve's attack, the composite system should satisfy:
\begin{equation}
\begin{aligned}
&U_FU_E(|0,0\rangle_{TE})=U_F(|0,E_0\rangle)=|0,F_0\rangle,\\
&U_FU_E(|1,0\rangle_{TE})=U_F(|1,E_3\rangle)=|1,F_3\rangle.
\end{aligned}
\end{equation}

\textbf{\emph{(2) Eve performs $(U_E,U_F)$ on the Bell states}}

Now, let us demonstrate the attack $(U_E,U_F)$ on the Bell state $|\phi^+\rangle$. We consider the worst case that Eve attacks both channels (i.e., TP to Alice, and TP to Bob) simultaneously. In more detail, Eve first attacks $U^{A}_E$ and $U^{B}_{E}$ on the qubits $P^{i}_{A}$ and $P^{i}_{B}$, which belong to the same Bell state. After Eve's first attack, the composite system is described as:
\begin{equation}
\begin{aligned}
&U^{A}_{E}U^{B}_{E}\left[\frac{1}{\sqrt 2}(|00\rangle+|11\rangle)_{AB}\otimes|00\rangle_{E_A E_B}\right]\\
&\quad =\frac{1}{\sqrt 2}(\alpha|0,E_0\rangle+\beta|1,E_1\rangle)_{AE_A}\otimes(\alpha|0,E_0\rangle+\beta|1,E_1\rangle)_{BE_B}\\
&\quad +\frac{1}{\sqrt 2}(\nu|0,E_2\rangle+\mu|1,E_3\rangle)_{AE_A}\otimes(\nu|0,E_2\rangle+\mu|1,E_3\rangle)_{BE_B},
\end{aligned}
\end{equation}
where the subscripts $E_A$ and $E_B$ denote the Eve's probe entangled on $P^{i}_{A}$ and $P^{i}_{B}$, respectively. $U^{A}_E$ and $U^{B}_{E}$ denote the unitary operators acting on the qubits $P^{i}_{A}$ and $P^{i}_{B}$.

According to the Eq. (9), the Eq. (12) can be evolved into
\begin{equation}
\begin{aligned}
&U^{A}_{E}U^{B}_{E}\left[\frac{1}{\sqrt 2}(|00\rangle+|11\rangle)_{AB}\otimes|00\rangle_{E_A E_B}\right]\\
&\quad =\frac{1}{\sqrt 2}\left[|0,E_0\rangle_{AE_A}\otimes|0,E_0\rangle_{BE_B}+|1,E_3\rangle_{AE_A}\otimes|1,E_3\rangle_{BE_B}\right]\\
&\quad =\frac{1}{\sqrt 2}\left(|00,E_0E_0\rangle+|11,E_3E_3\rangle\right)_{ABE_AE_B}.
\end{aligned}
\end{equation}
After Alice's (Bob's) operations, Eve will perform $U^{A}_{F}$ and $U^{B}_{F}$ on the qubits from Alice and Bob to TP, where $U^{A}_F$ and $U^{B}_{F}$ denote the unitary operators acting on the qubits $P^{i}_{A}$ and $P^{i}_{B}$. Here, we focus on the qubits that \textbf{\emph{reflect}} by Alice and Bob directly. Based on Eq. (10), the system will change to
\begin{equation}
\begin{aligned}
&U^{A}_{F}U^{B}_{F}\left[  \frac{1}{\sqrt 2}\left(|00,E_0E_0\rangle+|11,E_3E_3\rangle\right)_{ABE_AE_B}  \right]\\
&\quad =\frac{1}{\sqrt 2}\left(|00,F_0F_0\rangle+|11,F_3F_3\rangle\right)_{ABE_AE_B}\\
&\quad =\frac{1}{2}\left[|\phi^+\rangle(|F_0F_0\rangle+|F_3F_3\rangle)+|\phi^-\rangle (|F_0F_0\rangle-|F_3F_3\rangle)
\right].
\end{aligned}
\end{equation}
In this case, if Eve wants to be undetected by Alice and Bob, TP's measurement result on the qubits $P^{i}_{A}$ and $P^{i}_{B}$ should always be $|\phi^+\rangle$. Therefore, we have
\begin{equation}
|F_0\rangle=|F_3\rangle=|F\rangle.
\end{equation}

From Eqs. (9), (10), (11), and (15), if $(U_E,U_F)$ attacks on the single particles whitout inducing errors, we have
\begin{equation}
\begin{aligned}
&U_FU_E(|0,0\rangle_{TE})=|0,F_0\rangle=|0,F\rangle,\\
&U_FU_E(|1,0\rangle_{TE})=|1,F_3\rangle=|1,F\rangle.
\end{aligned}
\end{equation}
Similary, if $(U_E,U_F)$ attacks on the Bell states whitout inducing errors, we have
\begin{equation}
\begin{aligned}
&U^{A}_{F}U^{B}_{F}U^{A}_{E}U^{B}_{E}\left(|\phi^+\rangle_{AB}\otimes|00\rangle_{E_A E_B}\right)\\
&\quad =\frac{1}{2}|\phi^+\rangle_{AB}(|F_0F_0\rangle+|F_3F_3\rangle)_{E_A E_B}\\
&\quad =|\phi^+\rangle_{AB}\otimes|FF\rangle_{E_A E_B},
\end{aligned}
\end{equation}
where $|F\rangle_{E_A}$ and $|F\rangle_{E_B}$ denote Eve's probes on the qubits $P^{i}_{A}$ and $P^{i}_{B}$. It is not hard to see from Eqs. (16-17), Eve's probes are all in the same state of $|F\rangle$ (i.e., her probes are independent of the target qubit).

In summary, for Eve not inducing errors in steps 5 and 6, Eve's probe should be independent of the target qubit. That is to say, no matter what state the target qubit is in, Eve can only get the same result from her probe. Therefore, our protocol can withstand the entangle-measure attack.

\subsubsection{The Double-CNOT attack}
Semi-quantum protocol is the two-way protocol, so it susceptible to the Double-CNOT attack. Let us assume that Eve employs the Double-CNOT attack
to eavesdrop on Alice's secret information beacuse of the same role that Alice and Bob play. The detailed attack steps are as follows.  

Eve firstly intercepts the qubit from TP to Alice. Then, she performs the first attack $U_{CNOT}=|00\rangle \langle00|+|01\rangle \langle01|+|10\rangle \langle10|+|11\rangle \langle11|$, where the intercepted qubit is the control bit and her ancillary particle $\left|0\right\rangle_E$ is the target bit. After that, Eve intercepts the qubit from Alice to TP and executes the second $U_{CNOT}$ on the intercepted qubit and her ancillary particle again. Finally, Eve measures her ancillary particle to obtain some useful information.

However, Eve's Double-CNOT attack cannot succeed in our protocol. In the following, we analyze the effect of Eve's Double-CNOT attack on single particles and Bell states, respectively. Concretely speacking, for the single particles $|0\rangle$ and $|1\rangle$, after the first CNOT attack, the composite system will be:
\begin{equation}
U_{CNOT} \big(|0\rangle_A |0\rangle_E \big) =|00\rangle_{AE},
\end{equation}
\begin{equation}
U_{CNOT} \big(|1\rangle_A |0\rangle_E \big) =|11\rangle_{AE},
\end{equation}
where the subscripts $A$ and $E$ represent the target qubit and Eve's ancillary particle, respectively. Then, Alice performs the \textbf{\emph{measure}} operation or the \textbf{\emph{reflect}} operation on the received qubits. Note that, for the single particle, Alice's operation does not change its state. Subsequently, Eve performs the second CNOT attack on the qubit from Alice to TP and her ancillary particle. Thus, the composite system changes to
\begin{equation}
U_{CNOT} \big(|0\rangle_A |0\rangle_E \big) =|00\rangle_{AE},
\end{equation}
\begin{equation}
U_{CNOT} \big(|1\rangle_A |1\rangle_E \big) =|10\rangle_{AE}.
\end{equation}
From Eqs. (20-21), it easy to obtain that Eve's ancillary particle is always in the state of $|0\rangle$, no matter what the state of the target particle is. Therefore, Eve's attack for the single-particle is invalid.

Let us analyze the Bell state situation. After the first CNOT attack, the composite system will be:
\begin{equation}
\begin{aligned}
&U_{CNOT} \left[\frac{1}{\sqrt 2}(|00\rangle_{AB}+|11\rangle_{AB})|0\rangle_E \right]\\
&\quad =\frac{1}{\sqrt 2}(|000\rangle_{ABE}+|111\rangle_{ABE}),
\end{aligned}
\end{equation}
where $U_{CNOT}$ is perfromed on the qubits $A$ and $E$. If Alice chooses the \textbf{\emph{measure}} operation and obtains the measurement result $|0\rangle$ ($|1\rangle$), then after Eve's the second CNOT attack, the system becomes
\begin{equation}
\begin{aligned}
&U_{CNOT} |000\rangle_{ABE}= |000\rangle_{ABE},\quad  \text{if Alice obtains }|0\rangle \\
&U_{CNOT} |111\rangle_{ABE}= |110\rangle_{ABE}, \quad  \text{if Alice obtains }|1\rangle .
\end{aligned}
\end{equation}
If Alice chooses the \textbf{\emph{reflect}} operation, then after Eve's the second CNOT attack, the system becomes
\begin{equation}
\begin{aligned}
&U_{CNOT} \left[\frac{1}{\sqrt 2}(|000\rangle_{ABE}+|111\rangle_{ABE})\right]\\
&\quad =\frac{1}{\sqrt 2}(|00\rangle_{AB}+|11\rangle_{AB})|0\rangle_E.
\end{aligned}
\end{equation}
From Eqs. (23-24), it easy to obtain that Eve's ancillary particle is always in the state of $|0\rangle$. That is to say, Eve cannot obtain some useful information from her ancillary particle. Therefore, Eve's attack for the Bell state is invalid.

 \subsubsection{The Trojan horse attack}
In our protocol, each qubit is transmitted twice. To extract Alice's and Bob's secret, Eve may perform Trojan horse attacks \cite{39,40,41} which mainly include the delay-photon attack and the invisible photon attack. Fortunately, by equipping with wavelength filters and photon number splitters \cite{42,43}, the Trojan horse attacks can be effectively defended. As a result, our protocol can withstand the Trojan horse attacks.

\subsection{Internal attack}
Compared with external attackers, the attack from the adversarial users and TP may pose a greater threat to the security of the protocol. In more detail, two scenarios should be considered: one is that an adversarial user attempts to eavesdrop on the secret from another, the other is that TP wants to obtain the secret from the classical users.

\subsubsection{The attack from one adversarial user}
In the proposed protocol, Alice and Bob play the same role, either of whom may be adversarial. Without loss of generality, we suppose Alice is adversarial and she wants to obtain Bob's secret information. 

In our protocol, there are no qubits transmitted between Alice and Bob. That is to say, Alice is thoroughly independent of Bob. Therefore, to obtain the secret of Bob, Alice needs to intercept the qubits between Bob and TP. Unfortunately, if Alice intercepts Bob's qubits and launches the attacks mentioned above (e.g., intercept-resend attack, entangle-measure attack), she is the same as an external eavesdropper, Eve, and has no advantage. Alice essentially acts as an external eavesdropper. Based on the above analysis of the external attack, Bob and TP will detect Alice's attacks with a non-zero probability. 

In addition, Alice can obtain the secure key sequences $K_{AB}=[k^{1}_{AB},k^{2}_{AB},\dots,k^{n}_{AB}]$ and $K_{TA}=[k^{1}_{TA},k^{2}_{TA},\dots,k^{n}_{TA}]$ and the comparison result.  However, it is still helpless for her to get Bob's secret. Because Bob's secret is encrypted with $K_{TB}$ and $K_{AB}$, while Alice knows nothing about $K_{TB}$. The results show that one adversarial user cannot obtain other parties’ secret information.

\subsubsection{The attack from adversarial TP}
In our protocol, TP is assumed to be adversarial and has full quantum capability. More importantly, all quantum resources and complex quantum state measurements can only be accomplished by TP. In fact, TP' s attacks pose the greatest threat to protocol security because she can take all possible attacks to steal the useful information, including preparing fake quantum states. Recall the steps of our protocol, Alice's (Bob's) secret information is encrypted by $K_{AB}$ and $K_{TA}$ ($K_{AB}$ and $K_{TB}$), where the value of $K_{AB}$ is unknown to TP. To obtain the secret of Alice (Bob), TP must get $K_{AB}$.  There are two kinds of attacks that TP may use to obtain $K_{AB}$. 

For the first kind of attack, TP may measure the qubits $P^{i}_{A}$ and $P^{i}_{B}$ in $Z$-basis instead of Bell-basis, and then TP announces  her measurement result in the form of $|\phi^+\rangle$ or $|\phi^-\rangle$ randomly. In this way, TP can obtain Alice and Bob's measurement results on the qubits that they choose the \textbf{\emph{measure}} operation. However, this kind of attack will destroy the state of $|\phi^+\rangle$. For the qubits that Alice and Bob both choose the \textbf{\emph{reflect}} operation, TP will have a $\frac{1}{2}$ probability to publish the wrong measurement result $|\phi^-\rangle$. Thus, this kind of attack will be detected by Alice and Bob.

For the second kind of attack, TP may prepare fake particles with $Z$-basis instead of the Bell state $|\phi^+\rangle$, and then she sends these fake particles to Alice and Bob. After that, TP can measure the qubits from Alice and Bob with $Z$-basis to obtain Alice and Bob's measurement results. Following the protocol steps, TP is required to post the measurement result in the form of Bell state. Since TP prepares dummy particles, she can only publish the measurement results randomly as $|\phi^+\rangle$ or $|\phi^-\rangle$. Therefore, this kind of attack will be detected by Alice and Bob in step 5.

In addition, TP can obtain $K_{TA}$, $K_{TB}$, $Q_A$, $Q_B$ and the comparison result, in the course of the protocol. However, it is still helpless for TP to obtain Alice and Bob's secret, since she knows nothing about $K_{AB}$. The results show that the adversarial TP cannot obtain the classical users' secret information.

\section{Generalization to other semi-quantum protocols }

In the proposed protocol, Alice, Bob, and TP can establish the secure key sequence with each other. In more detail, Alice and Bob share a key sequence
$K_{AB}=[k^{1}_{AB},k^{2}_{AB},\dots,k^{n}_{AB}]$; Alice and TP share a key sequence $K_{TA}=[k^{1}_{TA},k^{2}_{TA},\dots,k^{n}_{TA}]$; Bob and TP have a key sequence $K_{TB}=[k^{1}_{TB},k^{2}_{TB},\dots,k^{n}_{TB}]$. Following this fact, our protocol can be applied to other semi-quantum cryptography protocols, such as semi-quantum key agreement and semi-quantum secure multi-party computation protocols.


\subsection{Semi-quantum key agreement protocol }

The goal of semi-quantum key agreement (SQKA) \cite{44,45,46,47} is to achieve the same contribution of all participants to the final shared key, where the capabilities of the participants are different and only one user is fully quantum capable, while other users' quantum capabilities are limited.

Here, we use three-party SQKA protocol as an example for illustration. Suppose that TP has full quantum capability, while Alice and Bob are two classical users with limited quantum power. They has a secret bit strings $m_A$, $m_B$ and $m_T$, respectively. That is
\begin{equation}
\begin{aligned}
&m_A=\{m^{1}_{A},m^{2}_{A},\dots,m^{n}_{A}\},\\
&m_B=\{m^{1}_{B},m^{2}_{B},\dots,m^{n}_{B}\},\\
&m_T=\{m^{1}_{T},m^{2}_{T},\dots,m^{n}_{T}\}.
\end{aligned}
\end{equation}
In this protocol, TP, Alice and Bob want to establish a secret key $K=m_A \oplus m_B\oplus m_T$, where all three parties contribute equally to construct the key. 

\subsubsection{The detailed steps of SQKA protocol}
In the following, we describe the SQKA protocol based on the previously proposed SQPC protocol. For simplicity, we would only like to introduce necessary steps that differ from the SQPC protocol above, while others are the same as those described in Sect. 2. 

\textbf{Step 1$^{SQKA}$ $\sim$ Step 6$^{SQKA}$:} These steps are the same as the SQPC protocol described in Sect. 2. After that, Alice, Bob, and TP can establish the secure key sequences with each other, denoted as $K_{AB}$, $K_{TA}$ and $K_{TB}$, respectively.

\textbf{Step 7$^{SQKA}$:} Alice, Bob and TP encrypt their secret bit stings $m_A$, $m_B$ and $m_T$ with $K_{AB}$, $K_{TA}$ and $K_{TB}$. More exactly, Alice uses $K_{AB}$ and $K_{TA}$ to encrypt $m_A$ as:
\begin{equation}
\begin{aligned}
&Q_{A\rightarrow B}=[m^{1}_{A}\oplus K^{1}_{AB},m^{2}_{A}\oplus K^{2}_{AB},\dots,m^{n}_{A}\oplus K^{n}_{AB}],\\
&Q_{A\rightarrow T}=[m^{1}_{A}\oplus K^{1}_{TA},m^{2}_{A}\oplus K^{2}_{TA},\dots,m^{n}_{A}\oplus K^{n}_{TA}],\\
\end{aligned}
\end{equation}
Bob uses $K_{AB}$ and $K_{TB}$ to encrypt $m_B$ as:
\begin{equation}
\begin{aligned}
&Q_{B\rightarrow A}=[m^{1}_{B}\oplus K^{1}_{AB},m^{2}_{B}\oplus K^{2}_{AB},\dots,m^{n}_{B}\oplus K^{n}_{AB}],\\
&Q_{B\rightarrow T}=[m^{1}_{B}\oplus K^{1}_{TB},m^{2}_{B}\oplus K^{2}_{TB},\dots,m^{n}_{B}\oplus K^{n}_{TB}],\\
\end{aligned}
\end{equation}
TP uses $K_{TA}$ and $K_{TB}$ to encrypt $m_T$ as:
\begin{equation}
\begin{aligned}
&Q_{T\rightarrow A}=[m^{1}_{T}\oplus K^{1}_{TA},m^{2}_{T}\oplus K^{2}_{TA},\dots,m^{n}_{T}\oplus K^{n}_{TA}],\\
&Q_{T\rightarrow B}=[m^{1}_{T}\oplus K^{1}_{TB},m^{2}_{T}\oplus K^{2}_{TB},\dots,m^{n}_{T}\oplus K^{n}_{TB}].\\
\end{aligned}
\end{equation}

\textbf{Step 8$^{SQKA}$:} Each of Alice, Bob and TP calculates the hash value of the corresponding encrypted messages and announces the result to the other two parties. The relationship of the encrypted messages and their hash values are shown in Table 1. Here $h(\cdot)$ is some one-way hash function. Note that this step is used to avoid information leaking and tampering due to the asynchronous release of information.
\begin{table}[htp]
\centering 
\tabcolsep 10pt
\caption{The relationship of the encrypted messages and their hash values. }
\label{tab:1}       
\begin{tabular}{ccc}
\hline
    The encrypted   &  Its hash value  & The recipient of   \\
         message    &   $h(\cdot)$     &  the hash value\\             
  \hline
  \noalign{\smallskip}
     $Q_{A\rightarrow B}$   &  $h(Q_{A\rightarrow B})$  &  Bob  \\
      \noalign{\smallskip}
     $Q_{A\rightarrow T}$   &  $h(Q_{A\rightarrow T})$  & TP  \\
      \noalign{\smallskip}
     $Q_{B\rightarrow A}$   &  $h(Q_{B\rightarrow A})$  & Alice  \\
      \noalign{\smallskip}
     $Q_{B\rightarrow T}$   &  $h(Q_{B\rightarrow T})$  & TP  \\
      \noalign{\smallskip}
     $Q_{T\rightarrow A}$   &  $h(Q_{T\rightarrow A})$  & Alice  \\
      \noalign{\smallskip}
     $Q_{T\rightarrow B}$   &  $h(Q_{T\rightarrow B})$  & Bob  \\
      \noalign{\smallskip}
\hline
\noalign{\smallskip}
\end{tabular}
\end{table}
%

\textbf{Step 9$^{SQKA}$:} After that, Alice, Bob, and TP publish their encrypted messages to the other two parties. Alice calculates the hash values of $Q_{B\rightarrow A}$ and $Q_{T\rightarrow A}$ to obtain the results $h^{\prime}(Q_{A\rightarrow B})$ and $h^{\prime}(Q_{A\rightarrow T})$. If 
$h^{\prime}(Q_{A\rightarrow B})=h(Q_{A\rightarrow B})$ and $h^{\prime}(Q_{A\rightarrow T})=h(Q_{A\rightarrow T})$, Alice will accept them. Then, Alice can decrypt $Q_{B\rightarrow A}$ and $Q_{T\rightarrow A}$ with $K_{AB}$ and $K_{TA}$ to obtain the secret keys $m_B$ and $m_T$. Using the same procedure, Bob and TP can also obtain $m_A$, $m_T$ and $m_A$, $m_B$, respectively.

\textbf{Step 10$^{SQKA}$:} Each of Alice, Bob, and TP has the secret keys $m_A$, $m_B$ and $m_T$. Thus, they can calculate the final key as 
$K=m_A \oplus m_B \oplus m_T$.

\subsubsection{Fairness of the proposed SQKA protocol}
In Section 3, we provide a detailed security analysis of the SQPC protocol, and since our proposed SQKA protocol is based on the SQPC protocol, the security analysis is similar. That is, Alice, Bob, and TP can establish the secure key sequences with each other. Unlike SQPC protocol, SQKA requires all involved parties equally contribute to the final shared key. Therefore, we focus on analyzing the fairness of each party's contribution to the key.
 
Without loss of generality, we assume that Alice wants to determine the shared key to be $K^*_{A}=[m^{*1}_{A},m^{*2}_{A},\dots,m^{*n}_{A}]$ alone. To achieve this goal, Alice needs to obtain $M_B$ and $M_T$. In our protocol, Alice has the opportunity to obtain $m_B$ and $m_T$ from Bob and TP announcing $Q_{B\rightarrow A}$ and $Q_{T\rightarrow A}$ only in step 9. After that, Alice calculates 
\begin{equation}
\begin{aligned}
&Q^{*}_{A\rightarrow B}=K^{*}_{A}\oplus m_{B}\oplus m_{T}\oplus K_{AB}\\
&Q^{*}_{A\rightarrow T}=K^{*}_{A}\oplus m_{B}\oplus m_{T}\oplus K_{TA},\\
\end{aligned}
\end{equation}
and Alice then publishes $Q^{*}_{A\rightarrow B}$ and $Q^{*}_{A\rightarrow T}$ to Bob and TP, respectively. As a result, Bob and TP can obtain the final key $K^{*}_{A}$ by computing $(K^{*}_{A}\oplus m_{B}\oplus m_{T})\oplus m_{B}\oplus m_{T}=K^{*}_{A}$. Unfortunately, Alice's behavior will be detected by Bob and TP. Because $Q^{*}_{A\rightarrow B}$ and $Q^{*}_{A\rightarrow T}$ can be accepted by Bob and TP only if they satisfy: $h^{\prime}(Q^{*}_{A\rightarrow B})=h(Q_{A\rightarrow B})$ and $h^{\prime}(Q^{*}_{A\rightarrow T})=h(Q_{A\rightarrow T})$. Obviously, they are not equal, so Alice's cheating behavior cannot succeed.

If two dishonest parties conspire to perform deceptions similar to those described above, it is inevitable that their deceptions will also be discovered by third parties due to the use of hash functions.

The results show that our protocol can guarantee all parties equally contribute to the final shared key.

\subsection{Semi-quantum secure multi-party computation protocols }
Quantum secure multi-party computation is an essential topic in quantum cryptography, which is to compute a function with private inputs from different parties in a distributed network and without revealing the true content of each private input. Quantum private comparison, quantum summation, quantum anonymous ranking and so on are branches of quantum secure multi-party computation. Inspired by the ideas of Refs. \cite{48,49}, in this part, we will show that our SQPC protocol can be generalized to semi-quantum summation and semi-quantum anonymous ranking protocols. 

For simplicity, we would only like to introduce necessary steps that differ from the SQPC protocol above, while others are the same as those described in Sect. 2.  As for the security, the extended protocols are all based on SQPC protocol, so their security analysis is similar and can be obtained by the same method. Therefore, to avoid duplication, we omitted the security analysis procedure.

\subsubsection{Semi-quantum summation protocol}

The purpose of quantum summation is to obtain the correct summation result without revealing the secret information of the private holder.
According to the definition of the quantum summation protocol, it should follow the rules given below:
\begin{itemize}
\item [\textbf{1.}] \emph{Corretness:} The summation result should be correct.
\item [\textbf{2.}] \emph{Security:} Users' private integer can not be leaked out to others without being detected.
\item [\textbf{3.}] \emph{Privacy:} Each user's private integer shoule be kept secret from others.
\end{itemize}

Semi-quantum summation (SQS) \cite{50,51} reduces the quantum capability of some users on top of quantum summation, thus reducing the burden of expensive quantum resources. Here, we use two-party semi-quantum summation protocol as an example for illustration (The multi-party case can be easily obtained by analogy).

Same as the SQPC protocol setting, Alice and Bob have the limited quantum capability, while TP has full quantum power and may be adversarial.
Alice and Bob have the private integer $MA$ and $MB$, respectively, where $MA,MB\in \mathbb{Z}$. They want to calculate the summation $MA+MB$ without revealing their secret information, through the help of TP. The detailed protocol steps are as follows.

\textbf{Step 1$^{SQS}$ $\sim$ Step 6$^{SQS}$:} These steps are the same as the SQPC protocol described in Sect. 2. After that, Alice, Bob, and TP can establish the secure key sequences with each other, denoted as 
\begin{equation}
\begin{aligned}
&K_{AB}=[k^{1}_{AB},k^{2}_{AB},\dots,k^{n}_{AB}],\\
&K_{TA}=[k^{1}_{TA},k^{2}_{TA},\dots,k^{n}_{TA}],\\
&K_{TB}=[k^{1}_{TB},k^{2}_{TB},\dots,k^{n}_{TB}],
\end{aligned}
\end{equation}
where $k^{j}_{AB},K^{j}_{TA}, K^{i}_{TB}\in \{0,1\}$, and $j=1,2,\dots,n$.

\textbf{Step 7$^{SQS}$:} Alice, Bob and TP convert the binary sequences $K_{AB}$, $K_{TA}$ and $K_{TB}$ to integer form. In particular, we use 
$KAB=\sum^{n}_{j=1}k^{j}_{AB}2^{j-1}$,  $KTA=\sum^{n}_{j=1}k^{j}_{TA}2^{j-1}$, and $KTB=\sum^{n}_{j=1}k^{j}_{TB}2^{j-1}$ to represent the integer forms of $K_{AB}$, $K_{TA}$ and $K_{TB}$, respectively.

\textbf{Step 8$^{SQS}$:} Alice uses $KAB$ and $KTA$ to encrypt her secret $MA$ as $QA=KAB+KTA+MA$. Similary, Bob uses $KAB$ and $KTB$ to encrypt her secret $MB$ as $QB=KAB+KTB+MB$. After that, they publish $QA$ and $QB$ to TP, and then TP calculates 
\begin{equation}
\begin{aligned}
&RT=QA+QB-KTA-KTB\\
&\quad =MA+MB+KAB+KAB.
\end{aligned}
\end{equation}
Finally, TP publishes $RT$ to Alice and Bob.

\textbf {Step 9$^{SQS}$:} Alice and Bob calculate $R=RT-KAB-KAB$ to obtain the final summation result $MA+MB$. In contrast, TP can not get the final summation result.

It is easy to see that the proposed protocol satisfies the requirements of the semi-quantum summation protocol. 
 
  \textbf{\emph{Corretness:}} In our protocol, Alice (Bob) uses $KAB$ and $KTA$ ($KAB$ and $KTB$) to encrypt her (his) secret $MA$ ($MB$). Then, TP calculates $RT=MA+MB+KAB+KAB$. Finally, Alice and Bob calculate $R=RT-KAB-KAB$ to obtain the final summation result $MA+MB$. Obviously, Alice and Bob can get the correct result.
  
  \textbf{\emph{Security:}} The security of this SQS protocol is based on the fact that a secure key relationship can be established between Alice, Bob, and TP. In the previous security analysis, we have shown that Alice, bob, and TP can establish a secure key relationship. Therefore, our protocol is secure. That is, users' private integer can not be leaked out to others without being detected.
  
  \textbf{\emph{Privacy:}} The user's secret information is encrypted with $K_{AB}$ and $K_{TA}$ ($K_{TB}$). Any eavesdropper has access to at most a portion of the encryption key. Therefore, each user's private integer is kept secret from others.

It can be concluded that the proposed SQPC protocol can be generalized to semi-quantum summation protocol.

\subsubsection{Semi-quantum anonymous ranking protocol}

Quantum anonymous ranking (QAR)\cite{52,53,54} aims to anonymously rank the data of multiple users, where each user should know the positions of his numbers in ascending (or descending) sequence of the ranked numbers. The QRA protocol must comply with the following rules:

\begin{itemize}
\item [(1)] \emph{Corretness:} Each user can correctly get the ranking (ascending or descending) of her data.
\item [(2)] \emph{Anonymity:} For each user, no one else has access to information about her data ranking.
\item [(3)] \emph{Privacy:} The data values for each user should be kept confidential to all others.
\end{itemize}
It should be noted that in a two-party scenario, there is no secure ranking protocol, because one user can directly obtain the data ranking of another user based on his own data ranking. Therefore, the number of users involved in the ranking protocol should be greater than $2$.

Inspired by the ideas of Ref. \cite{52}, we combine semi-quantum ideas into quantum anonymous ranking and propose a semi-quantum anonymous ranking protocol (SQAR). Similar to the previous ones, the proposed semi-quantum anonymous sorting protocol is generalized from the multi-party SQPC protocol (which is given in Sect.  2.2).

Assume that there are $L (L>2)$ ``classical'' users labeled $C_1,C_2,\cdots,C_L$, where $C_l (l=1,2,\dots,L)$ has a secret number $m_l\in\{1,2,\dots,N\}$ and $N\in \mathbb{Z^+}$. The purpose of our protocol is to rank all the data of ``classical'' users with the help of TP, whose quantum power is unlimited.

\textbf{Step 1$^{SQAR}$ $\sim$ Step 4$^{SQAR}$:} These stpes are the same as the \textbf{Step 1* $\sim$ Step 4*} described in Sect. 2.2.

\textbf{Step 5$^{SQAR}$:} $C_1,C_2,\cdots,C_L$ discuss eavesdropping and TP's honesty in this step. Here, we focus on two case:

\begin{itemize}
\item [\textbf{1)}] If all $C_1,C_2,\cdots,C_L$ choose the \textbf{\emph{reflect}} operation on the qubits $P^{i}_{C_1},P^{i}_{C_2},\dots,P^{i}_{C_L}$, TP should always get $|\Psi^+\rangle$. This case is used for checking TP's honesty and eavesdropping. When the error rate of this case surpasses the threshold, the protocol ends.

\item [\textbf{2)}] For the arbitrary two users $C_l$ and $C_g$ ($l,g=1,2,\dots,L$ and $l\neq g$), if they choose \textbf{\emph{measure}} operation on the received qubits, while all other users choose \textbf{\emph{reflect}} operation, they will have the same measurement results recored as $K_{l,g}=[k^{1}_{l,g},k^{2}_{l,g},\dots,k^{n}_{l,g}]$. Note that all other users including TP know nothing about $K_{l,g}$, because TP performs $L$-particle GHZ state measurements, not $Z$-basis measurements, and all other ``classcial'' users choose \textbf{\emph{reflect}} operation instead of \textbf{\emph{measure}} operation.
\end{itemize}

\textbf{Step 6$^{SQAR}$:} This step is the same as the \textbf{Step 6*} described in Sect. 2.2.  After steps 1 to 6, TP can establish a secure key sequence with each ``classical'' user, represented as $K_{TC_l}=[k^{1}_{TC_l},k^{2}_{TC_l},\dots,k^{n}_{TC_l}]$. Then, arbitrary two ``classical'' users $C_l$ and $C_g$ can establish a secure key sequence $K_{l,g}=[k^{1}_{l,g},k^{2}_{l,g},\dots,k^{n}_{l,g}]$. Here, $k^{j}_{l,g} ,k^{j}_{TC_l}\in\{0,1\}$ and $j=1,2,\dots,n$.

\textbf{Step 7$^{SQAR}$:}  $C_l$ and TP convert the binary sequences $K_{l,g}$ and $K_{TC_l}$ to integer form. In particular, we use 
$K^{*}_{l,g}=\sum^{n}_{j=1}k^{j}_{l,g}2^{j-1}$ and  $K^{*}_{TC_l}=\sum^{n}_{j=1}k^{j}_{TC_l}2^{j-1}$ to represent the integer forms of $K_{l,g}$ and $K_{TC_l}$, respectively.

\textbf{Step 8$^{SQAR}$:} Each ``classical'' user generates a string of length $N$ as their sub-secret string. For $C_l$, her sub-secret string is $V_l$, where the $t$-th element $V^{t}_{l}$ is $K^{*}_{TC_l}-K^{*}_{l-1,l}+K^{*}_{l,l+1}$, $t=1,2,\dots,N$. It's important to point out that every element of $V_l$ is the same, and $K^{*}_{L,L+1}=K^{*}_{0,1}=K^{*}_{L,1}$.

\textbf{Step 9$^{SQAR}$:} Then $C_l$ encodes her data on the sub-secret sting $V_l$. Specifically, she replaces $m_l$-th element $V^{m_l}_{l}$ with $V^{m_l}_{l}+1$, while the ones on the other positions remain the same. After that, the new string is marked as $V^{\prime}_{l}$, and it will be sent to TP.

\textbf{Step 10$^{SQAR}$:} Based on the received information and $K^{*}_{TC_l}$ ($l=1,2,\dots,L$), TP calculates:
\begin{equation}
R^t= \sum^{L}_{l=1}V^{\prime t}_{l}-\sum^{L}_{l=1}K^{*}_{TC_l}.
\end{equation}
Then, TP publishes all the values $R^t$ for $t=1,2,\dots,N$. From the published information, each user has anonymous access to the ranking of their data. For example, $C_l$ can get the ranking of her data $m_l$ by computing
\begin{equation}
R^1+R^2+\dots+R^{m_l-1}+1,
\end{equation}
and nobody else knows that $m_l$ belongs to her. 

It is not difficult to see that the proposed protocol satisfies the correctness, anonymity and privacy. Let us show them in below. 

 \textbf{\emph{Corretness:}} For simplicity,  we use three-party semi-quantum anonymous ranking protocol as an example for illustration. Suppose $C_1$, $C_2$, and $C_3$ are thee ``classical'' users, and their secret data are $m_1=1$, $m_2=2$, and $m_3=3$. In our protocol, $C_1$, $C_2$, and $C_3$ respectively establish secret keys with TP, denoted as $K^{*}_{TC_1}$, $K^{*}_{TC_2}$, and $K^{*}_{TC_3}$. Moreover, $C_1$, $C_2$, and $C_3$ have also established secret keys between each other. The secret key relationships between all users are shown in Table 2.
 \begin{table}[htp]
\centering 
\tabcolsep 8pt
\caption{The secret key relationship between all users. }
\label{tab:1}       
\begin{tabular}{ccccc}
\hline
       &  $C_1$  & $C_2$ &  $C_3$ &  TP  \\        
  \hline
  \noalign{\smallskip}
   $C_1$ &  /  & $K^{*}_{1,2}$ & $K^{*}_{3,1}$ &  $K^{*}_{TC_1}$ \\
      \noalign{\smallskip}
   $C_2$ &   $K^{*}_{1,2}$    &  /  &  $K^{*}_{2,3}$   & $K^{*}_{TC_2}$  \\
      \noalign{\smallskip}
   $C_3$ &  $K^{*}_{3,1}$  &$K^{*}_{2,3}$  & /  &  $K^{*}_{TC_3}$  \\
      \noalign{\smallskip}
   TP &  $K^{*}_{TC_1}$  & $K^{*}_{TC_2}$ & $K^{*}_{TC_3}$  & / \\
      \noalign{\smallskip}
\hline
\noalign{\smallskip}
\end{tabular}
\end{table}
Then $C_1$, $C_2$, and $C_3$ generate a string of length $3$ as their sub-secret string, respectively. For $C_1$, her sub-secret string is $V_1$, where each element is $K^{*}_{TC_1}-K^{*}_{3,1}+K^{*}_{1,2}$. Afterwards $C_1$ encodes her data on the sub-secret sting $V_1$. Specifically, she replaces $m_l$-th (i.e., $1$-th) element $V^{1}_{1}$ with $V^{1}_{l}+1$, while $V^{2}_{1}$ and $V^{3}_{1}$ remain unchanged. $C_2$ and $C_3$ can get $V_2$ and $V_3$ in the same way.  After that, they publish $V^{\prime}_{1}$, $V^{\prime}_{2}$, and $V^{\prime}_{3}$ to TP, and TP calculates
\begin{equation}
\begin{aligned}
&R^1=\sum^{3}_{l=1}V^{\prime 1}_{l}-\sum^{3}_{l=1}K^{*}_{TC_l}=1,\\
&R^2=\sum^{3}_{l=1}V^{\prime 2}_{l}-\sum^{3}_{l=1}K^{*}_{TC_l}=1,\\
&R^3=\sum^{3}_{l=1}V^{\prime 3}_{l}-\sum^{3}_{l=1}K^{*}_{TC_l}=1.
\end{aligned}
\end{equation}
Then, TP announces $R^1$, $R^2$, and $R^3$ via a publich channel. According to their own secret data, $C_1$, $C_2$, and $C_3$ can get the ranking of their data by computing $R^1+R^2+\dots+R^{m_l-1}+1$. For example, $C_3$'s secret data is $3$, then she calculates $R^1+R^2+1=1+1+1=3$.  The results show that the proposed protocol can guarantee the correctness of the output.

 \textbf{\emph{Anonymity:}} In our protocol, each user's private data is kept secret from others. From the information published on the channel, the eavesdropper cannot deduce the user's private data ranking. If $C_l$ wants to obtain the ranking of her data, all she needs to do is to use her data to compute Eq. (33). The whole process is independent of other users. That is, the users can get the rankings of their data anonymously.

 \textbf{ \emph{Privacy:}} For $C_l$, her sub-secret string is encrypted with $K_{TC_l}$, $K^{*}_{l-1,l}$, and $K^{*}_{l,l+1}$. Any eavesdropper has access to at most a portion of the encryption key. In our protocol, TP performs only simple calculations and has no access to the user's private data. Moreover, eavesdroppers cannot obtain secret data from published messages. Therefore, each user's private data is kept secret from others.


\section{ Discussion }
\label{sec:5}

In this paper, we propose several different types of semi-quantum protocols. To better highlight the characteristics of the protocols, we compare the proposed protocols with their respective counterparts, separately. Before comparing the proposed protocols, we would like to point out that
 most of the current semi-quantum protocols (e.g., SQPC, SQS, SQKA) are two-party scenarios, and therefore, for the sake of fairness, our comparison process is conducted in a two-party scenario.

\subsection {Comparison of the SQPC protocol}
Performance of the semi-quantum private compariosn protocol can be characterized using qubit efficiency\cite{44}. Therefore, in the followings, we focus on the qubit efficiency of the protocol as a metric. Qubit efficiency is defined as $\eta=\frac{c}{q+b}$, where $c$, $q$, and $b$ are the number of shared classical bits, the number of consumed qubits, and the number of classical bits needed, respectively.

We first calculate the qubit efficiency of the proposed SQPC protocol.  In our protocol, Alice and Bob have $n$ secret bits, respectively, which means $c=n$. To implement the protocol, TP needs to generate $4n$ Bell states, and $8n$ single particles, while Alice and Bob need to prepare $4n$ qubits, respectively as a replacement for their measured particles. Thus, the number of consumed qubits is $24n$. Then, Alice and Bob need $2n$ bits to publish their encrypted messages, and TP needs 1 bit to publish the comparison result. That is, $b=2n+1$. Putting  everything together, the qubit efficiency of our protocol is $\frac{n}{26n+1}$. Similarly, the qubit efficiency of Refs. \cite{34,35,36,37,38} can be calculated, and the specific results are shown in Table 3. Pre-shared keys and scalability as two important indicators are also compared in Table 3. It’s clear from the table that our protocol has comparable advantages in the above three aspects. 

\begin{table}[htp]
\centering 
\tabcolsep 10pt
\caption{Comparison between our SQPC protocol and previous ones }
\label{tab:1}       
\begin{tabular}{cccc}
\hline
     &  Qubit efficiency  & Pre-shared keys & Scalability (Applicable   \\   
     &                    &                 & to multi-party scenarios)   \\      
  \hline
  \noalign{\smallskip}
    Ref. \cite{34}&  $\frac{n}{102n+1}$  & Yes & \ding{53}  \\
      \noalign{\smallskip}
    Ref. \cite{35} &   $\frac{n}{60n+1}$   &  Yes  &  \ding{53}     \\
      \noalign{\smallskip}
   Ref. \cite{36} &  $\frac{n}{52n+1}$  & No  & \ding{53}  \\
      \noalign{\smallskip}
   Ref. \cite{37} &  $\frac{n}{53n+1}$  & Yes  & \ding{53}  \\
      \noalign{\smallskip}
   Ref. \cite{38} &  $\frac{n}{18n+1}$  & No  & \checkmark \\
      \noalign{\smallskip}
   Our protocol &  $\frac{n}{26n+1}$  & No & \checkmark   \\
      \noalign{\smallskip}
\hline
\end{tabular}
\end{table}

\subsection {Comparison of the SQKA protocol }

For SQKA protocol, the qubit efficiency, number of users and scalability are important evaluation metrics. In the following, we compare the proposed SQKA protocol with similar ones \cite{44,45,46,47} in these three aspects.
 
We first calculate the qubit efficiency of the proposed SQKA protocol.  In this SQKA protocol, $24n$ qubits are consumed in order to implement the key negotiation of three users and finally generate a shared key of $n$ bits. As for the consumed classical bits, each of user needs to publish $n$-bits encrypted messages to the other users, along with the hash of the encrypted message (suppose the length of the hash value is $n$-bits). Thus, the total classical bits is $3*2n+3*2n=12n$. Putting  everything together, the qubit efficiency of our protocol is $\frac{1}{36}$.
Table 4 gives the detailed comparison results of our protocol with similar protocols. One can easily observe that the qubit efficiency of three-party schemes is less than that of two-party schemes. It is obvious that as the number of classical users increases, the quantum resources required will also increase. Compared with similar ones, our protocol has advantages in qubit efficiency and scalability.

\begin{table}[htp]
\centering 
\tabcolsep 10pt
\caption{Comparison between our SQKA protocol and similar ones }
\label{tab:1}       
\begin{tabular}{cccc}
\hline
     &  Qubit efficiency  & Number of users & Scalability (Applicable   \\   
     &                    &         & to multi-party scenarios)   \\      
  \hline
  \noalign{\smallskip}
    Ref. \cite{44}&  $\frac{1}{10}$  & Two & \ding{53}  \\
      \noalign{\smallskip}
    Ref. \cite{45} &   $\frac{1}{15}$   &  Two  &  \ding{53}     \\
      \noalign{\smallskip}
    Ref. \cite{46} &   $\frac{1}{48}$   &  Three  &  \ding{53}     \\
      \noalign{\smallskip}
    Ref. \cite{47} &  $\frac{1}{38}$  & Three  & \checkmark  \\
      \noalign{\smallskip}
    Our protocol &  $\frac{1}{36}$  & Three & \checkmark   \\
      \noalign{\smallskip}
\hline
\end{tabular}
\end{table}

\subsection {Comparison of the SQS and SQAR protocols}

There are few kinds of research on semi-quantum summation and semi-quantum anonymous ranking. As far as we know, only Zhang et al.\cite {50} and   Ye et al. \cite{51} have proposed semi-quantum summation protocols, and no one has yet to consider how to implement semi-quantum anonymous ranking.  In the following, we discuss the SQS protocol and SQAR protocol, separately.

Let us discuss the SQS protocol first. The previous SQS protocols were for binary numbers and could not be extended to multi-party scenarios. Our protocol, however, enables summation computation in integer form and can be extended to multiparty scenarios. Besides, in our SQS protocol and Ye et al.\cite{51}'s protocol, TP can not obtain the final summation result, while Zhang et al \cite{50}'s protocol cannot achieve this. Therefore, our protocol has a wider application than the previous two protocols. The detailed comparison results are list in Table 5.

\begin{table}[!htp]
\centering 
\tabcolsep 10pt
\caption{The comparison results of our SQS protocol and similar ones }
\label{tab:1}       
\begin{tabular}{cccc}
\hline
                        &  Ref. \cite{50}   & Ref. \cite{51} & Our protocol  \\ 
                        \noalign{\smallskip}
      \hline 
     \noalign{\smallskip}
    Quantum states      & Single particles & Two-qubit             & Entangled stats   \\
     used               &                  &  entangled states     & and single particles  \\
      \noalign{\smallskip}
    TP knows the        & Yes              &      No               &  No  \\
    result of summation &                  &                       &   \\
       \noalign{\smallskip}
     The data type for      & Binary number    &      Binary number    &  Integer  \\
      summation      &                  &                       &   \\
      \noalign{\smallskip}
   Scalability (Applicable  &   \ding{53}        &      \ding{53}               &   \checkmark  \\
  to multi-party scenarios)  &                  &                       &   \\
      \noalign{\smallskip}
\hline
\end{tabular}
\end{table}

Then, we discuss the proposed SQAR protocol. Compared with the previous QAR protocols \cite{52,53,54}, the most critical feature of the SQAR protocol is that it reduces the quantum capability of some participants, thus relieving the pressure caused by expensive quantum resources. Of course, this does not come without a cost. Due to the reduced quantum capabilities of users, semi-quantum protocols require more qubits and communication counts to ensure the protocol's security. Thus semi-quantum protocols are generally less efficient than full-quantum protocols. How to balance efficiency and quantum resources needs more attention in the future. But in any case, our proposed SQAR protocol provides a solution for quantum-limited users who want to rank their data anonymously.

\section{Conclusion }
\label{sec:6}

In this paper, we present a new SQPC protocol based on entangled states and single particles, where the classical users can use entangled states to establish secure keys with the help of TP, while the single particles can be used to establish secure keys between TP and classical users.
The proposed SQPC protocol does not require an additional SQKD protocol to pre-share the key between two classical users, which greatly improves the efficiency of the protocol. Then, based on multi-particle entangled states and single-particles our protocol can be easily extended to multi-party scenarios to accommodate multiple classical users who want to compare their privacy data. In addition, we generalize the proposed SQPC protocol to other semi-quantum protocols such as SQKA, SQS and SQAR. It is worthly pointing that we are the first to propose an SQAR protocol, and no one has considered the combination of semi-quantum and quantum anonymous ranking before. We also show that our protocols can withstand both the attacks from outside eavesdroppers and adversarial participants.

Many interesting future questions remain to be addressed. First, we have only generalized our protocols to SQKA, SQS and SQAR, and there are still more application scenarios to be discovered. Second, we only consider ideal environments; practical devices in semi-quantum environments are just starting to be realized\cite{55,56}, and applying some of these techniques to the protocol we present here is interesting.

\section*{Acknowledgments} 
This work was supported by the BUPT Excellent Ph.D Students Foundation under Grant CX2021117, the National Natural Science Foundation of China under Grant 92046001, 61962009, the Fundamental Research Funds for the Central Universities under Grant 2019XD-A02.

\end{document}